\documentclass[pdftex]{spie}  %>>> use for US letter paper

\usepackage[nooneline]{subfigure}
\usepackage{amsmath,amsfonts,amssymb}
\usepackage{graphicx}
\usepackage[colorlinks=true, allcolors=blue]{hyperref}
\usepackage{color}
\usepackage{authblk}

\title{Proton radiation damage tolerance of wide dynamic range SOI pixel detectors}

\author[a]{Shun Tsunomachi}
\author[a]{Takayoshi Kohmura}
\author[b]{Kouichi Hagino}
\author[a]{Masatoshi Kitajima}
\author[a]{\\Toshiki Doi}
\author[a]{Daiki Aoki}
\author[a]{Asuka Ohira}
\author[a]{Yasuyuki Shimizu}
\author[a]{Kaito Fujisawa}
\author[a]{\\Shizusa Yamazaki}
\author[a]{Yuusuke Uchida}
\author[a]{Makoto Shimizu}
\author[a]{Naoki Itoh}
\author[c]{Yasuo Arai}
\author[c]{\\Toshinobu Miyoshi}
\author[c]{Ryutaro Nishimura}
\author[d]{Takeshi Go Tsuru}
\author[e]{Ikuo Kurachi}

\affil[a]{Tokyo University of Science, School of Science and Technology, Department of Physics, 2641 Yamazaki, Noda, Chiba, Japan, 278-8510}
\affil[b]{Kanto Gakuin University, Research Advancement and Management Organization, 1-50-1 Mutsuura-higashi, Kanazawa-ku, Yokohama, Japan, 236-8501}
\affil[c]{High Energy Accelerator Research Organization (KEK), Institute of Particle and Nuclear Studies, 1-1 Oho, Tsukuba, Ibaraki, Japan, 305-0801}
\affil[d]{Kyoto University, Faculty of Science, Department of Physics, Kitashirakawa-Oiwakecho, Sakyo-ku, Kyoto, Japan, 606-8502}
\affil[e]{D\&S Inc., 774-3-213 Higashiasakawacho, Hachioji, Tokyo, Japan, 193-0834}

\begin{document} 
\maketitle

\begin{abstract}
We have been developing the SOI pixel detector ``INTPIX'' for space use and general purpose applications such as the residual stress measurement of a rail and high energy physics experiments. INTPIX is a monolithic pixel detector composed of a high-resistivity Si sensor, a $\mathrm{SiO_{2}}$ insulator, and CMOS pixel circuits utilizing Silicon-On-Insulator (SOI) technology. We have considered the possibility of using INTPIX to observe X-ray polarization in space. When the semiconductor detector is used in space, it is subject to radiation damage resulting from high-energy protons. Therefore, it is necessary to investigate whether INTPIX has high radiation tolerance for use in space. The INTPIX8 was irradiated with 6 MeV protons up to a total dose of 2 krad at HIMAC, National Institute of Quantum Science in Japan, and evaluated the degradation of the performance, such as energy resolution and non-uniformity of gain and readout noise between pixels. After 500 rad irradiation, which is the typical lifetime of an X-ray astronomy satellite, the degradation of energy resolution at 14.4~keV is less than 10$\%$, and the non-uniformity of readout noise and gain between pixels is constant within 0.1$\%$.
\end{abstract}


\section{INTRODUCTION}
CCDs have been used for about three decades because of their good image resolution as well as their moderate energy resolution and time resolution. However, they have drawbacks such as low quantum efficiency of hard X-ray due to a thin depletion layer,  difficulty in high-speed processing and degradation of Charge Transfer Efficiency (CTE) due to radiation damage, which degrades spectral performance. To overcome these drawbacks, we have been developing a Silicon-On-Insulator (SOI) pixel detector named \color{black}``SOIPIX.'' SOIPIX is composed of a CMOS circuit layer, a $\mathrm{SiO_{2}}$ insulator layer called BOX (Buried Oxide), and a Si sensor. SOIPIX has two features. First, a thick depletion layer of about 500$\mathrm{~\mu m}$ can be formed because Si with high resistivity can be selected for the sensor layer, and a high back bias voltage can be applied to the sensor layer due to the separation by the insulator layer. Second, because adjacent transistors are insulated from each other, the distance between transistors can be brought closer together, allowing transistors to be arranged in high density. 

There are several types of SOI, such as INTPIX, XRPIX, and SOFIT, depending on the purpose of use. XRPIX has been developed for space X-ray observations\cite{10.1117/12.2312098}, SOFIST has been developed for high energy physics experiments\cite{Yamada_2018}, and INTPIX has been developed for general use in ground experiments, such as stress measurement\cite{SASAKI2020164426}. XRPIX achieves a high time resolution of about $\sim10\mathrm{~\mu s}$ by installing a trigger function to pixels. For this reason, the pixel size of XRPIX is quite large at 36 $\mathrm{\mu m}$. SOFIST and INTPIX do not have a trigger function, so the pixel size is 20 $\mathrm{\mu m}$ and 16 $\mathrm{\mu m}$. \color{black}Therefore, INTPIX may also be used for space applications such as polarization measurements. For example, when a 200 keV gamma-ray is incoming, the maximum energy of its recoil electron is about 88 keV, and the range is about 65 $\mathrm{\mu m}$. If we detect the track and the energy deposit, we can estimate the degree of polarization of an incoming X-ray. We need a detector with a fine pixel size and good energy resolution to detect X-ray of a celestial object. A polarization measurement is difficult with XRPIX but possible with INTPIX with the aspect of the pixel size.
 
When Si semiconductor detectors such as CCD are used in space, CTE is known to degrade due to radiation damage. Radiation ionization also causes a change in the threshold voltage of CMOS circuits because a positive charge with low mobility is accumulated in the $\mathrm{SiO_{2}}$ layer. In addition, the interface level increase because the bond between Si and $\mathrm{SiO_{2}}$ is broken. The resulting increase in dark current degrades spectral performance. We are concerned about proton damage by the South Atlantic Anomaly (SAA) in low earth orbit. The absorbed dose from SAA is about 100 rad/year in a typical silicon detector. The typical lifetime of an X-ray astronomy satellite is about 5 years. The total amount of dose will be approximately 500 rad. Since we have not yet evaluated the degree of performance degradation of INTPIX for use in the space environment, we report the results of that evaluation in this paper. 

\section{EXPERIMENT OF PROTON IRRADIATION}
\begin{figure} [ht]
\begin{center}
\begin{tabular}{c} 
\includegraphics[height=7cm]{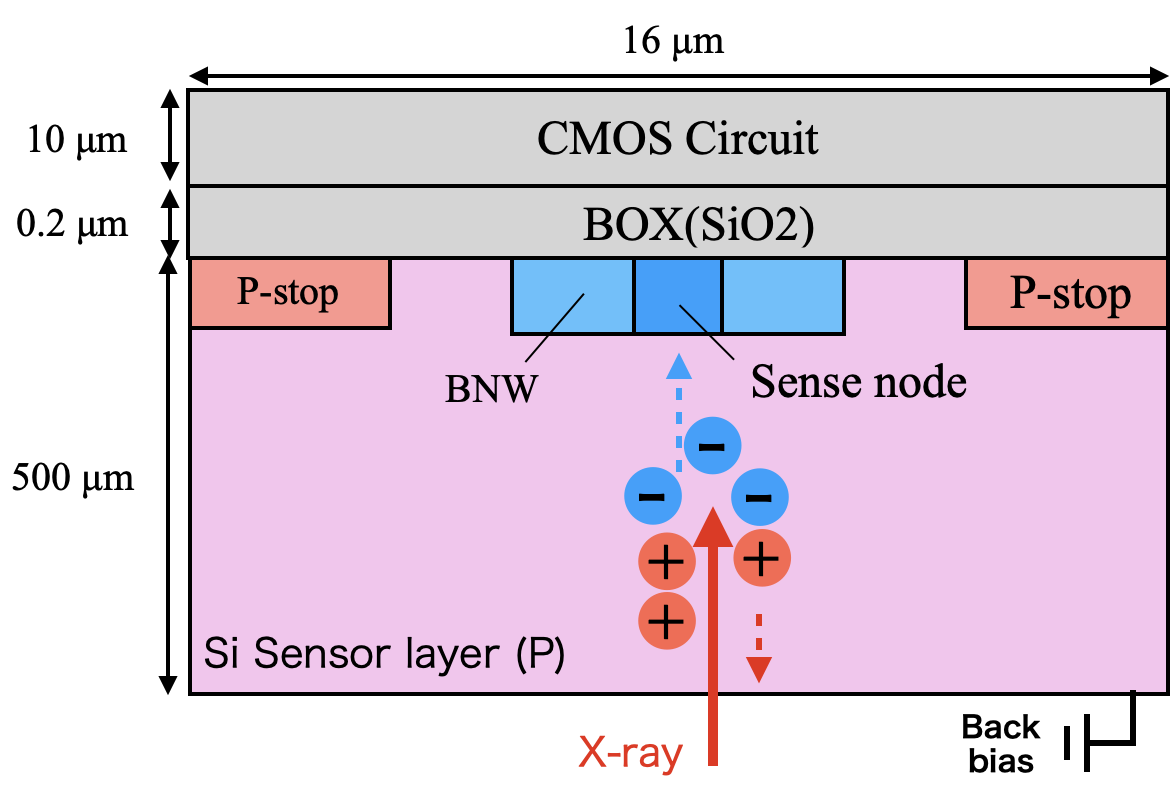}
\end{tabular}
\end{center}
\caption[example] 
{ \label{fig:INTPIX8_pix}
Schematic diagram of INTPIX8 pixel center}
\end{figure}

We irradiated a 6 MeV proton beam to INTPIX8 at HIMAC, National Institute of Quantum Science in Japan. A schematic of the cross-section of the pixel center is shown in Fig. \ref{fig:INTPIX8_pix}. Charges generated in the sensor layer are collected in the Sense Node and converted to voltage in CMOS circuits. The long horizontal BNW prevents high voltages (back bias) from being transmitted to CMOS circuits. The p-stop is introduced to create a lateral electric field to direct the charge toward the sense node.

\begin{figure} [ht]
\begin{center}
\begin{tabular}{c} 
\includegraphics[height=8cm]{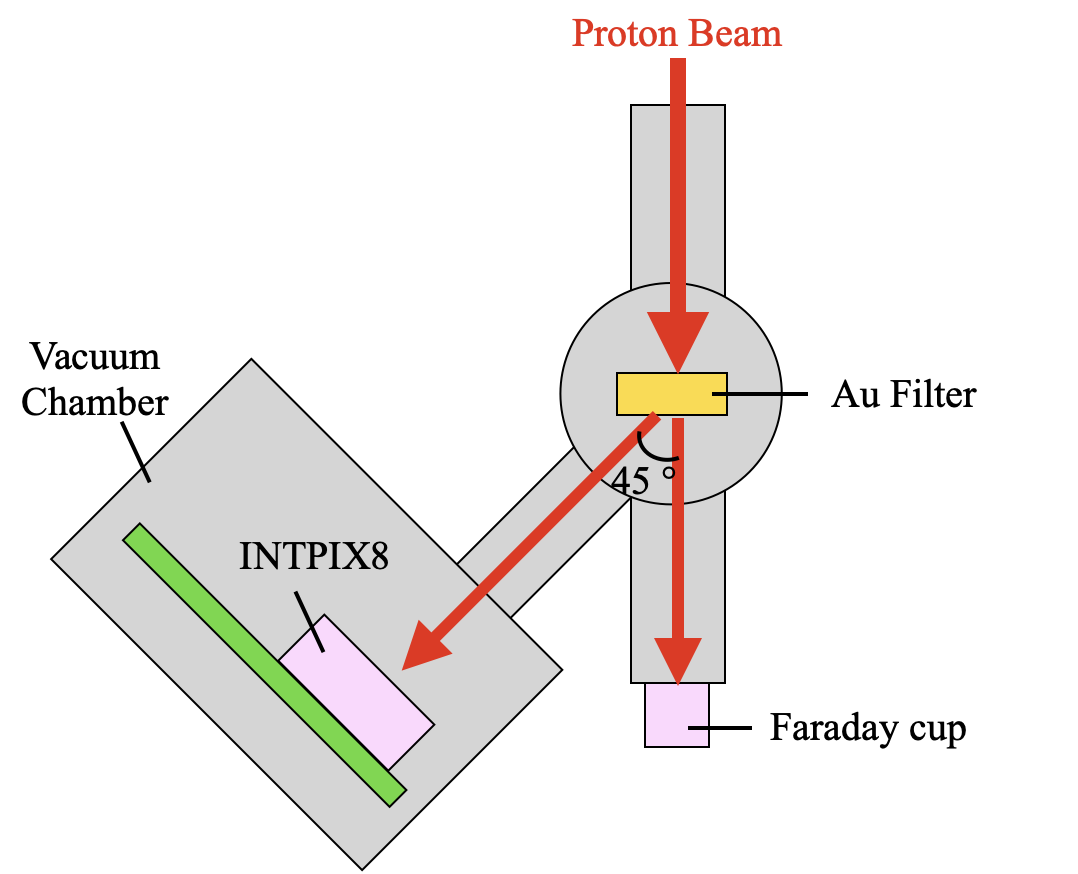}
\end{tabular}
\end{center}
\caption[example] 
{ \label{fig:Chamber}Setup of proton irradiation experiment}
\end{figure}

The experimental setup is shown in Fig. \ref{fig:Chamber}\cite{YARITA2019457}. The beam energy of 6 MeV is sufficiently high to evaluate the radiation damage due to the total ionizing effect because it uniformly ionizes in the BOX layer. We placed the Au filter to scatter the proton beam for 45 degrees. The Faraday cup (FC) was placed on the straight line of the beam. It can measure non-scattered protons and track changes in beam intensity. In order to determine proton beam intensity by using the FC, we need to calibrate. We replaced INTPIX8 with an avalanche photo-diode before the experiment and calibrated the FC with it. After that, we evaluated the amount of dose to the BOX layer of INTPIX8 with monitoring counts of the FC during proton irradiation. Spectral performance was evaluated at doses of 0, 500, 1000, and 2000 rad by irradiating X-rays from two radioisotopes, $^{57}$Co and $^{109}$Cd.

\section{Result}
%The performance of INTPIX8 before and after \color{red}proton irradiation \color{black}was evaluated in four aspects: dark current, readout noise, gain, and energy resolution.

For the performance evaluation of INTPIX8 after proton irradiation, we evaluated dark current, readout noise, gain, and energy resolution. By evaluating these, we can find how dark current and readout noise contribute to the degradation of spectral performance, such as gain and energy resolution.

\begin{figure} [ht]
\begin{center}
\begin{tabular}{c} 
\subfigure{%
\includegraphics[height=5cm]{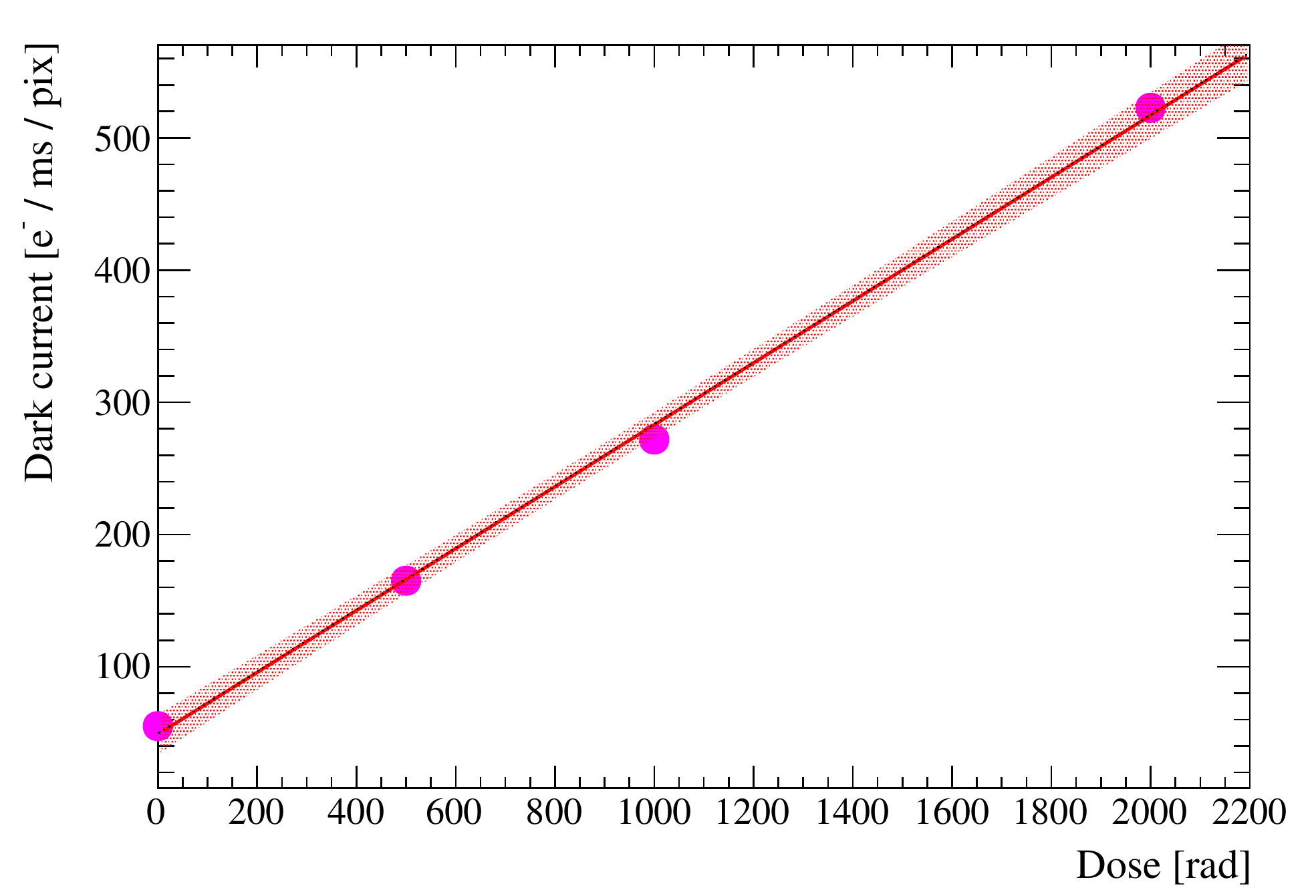}}%
\subfigure{%
\includegraphics[height=5cm]{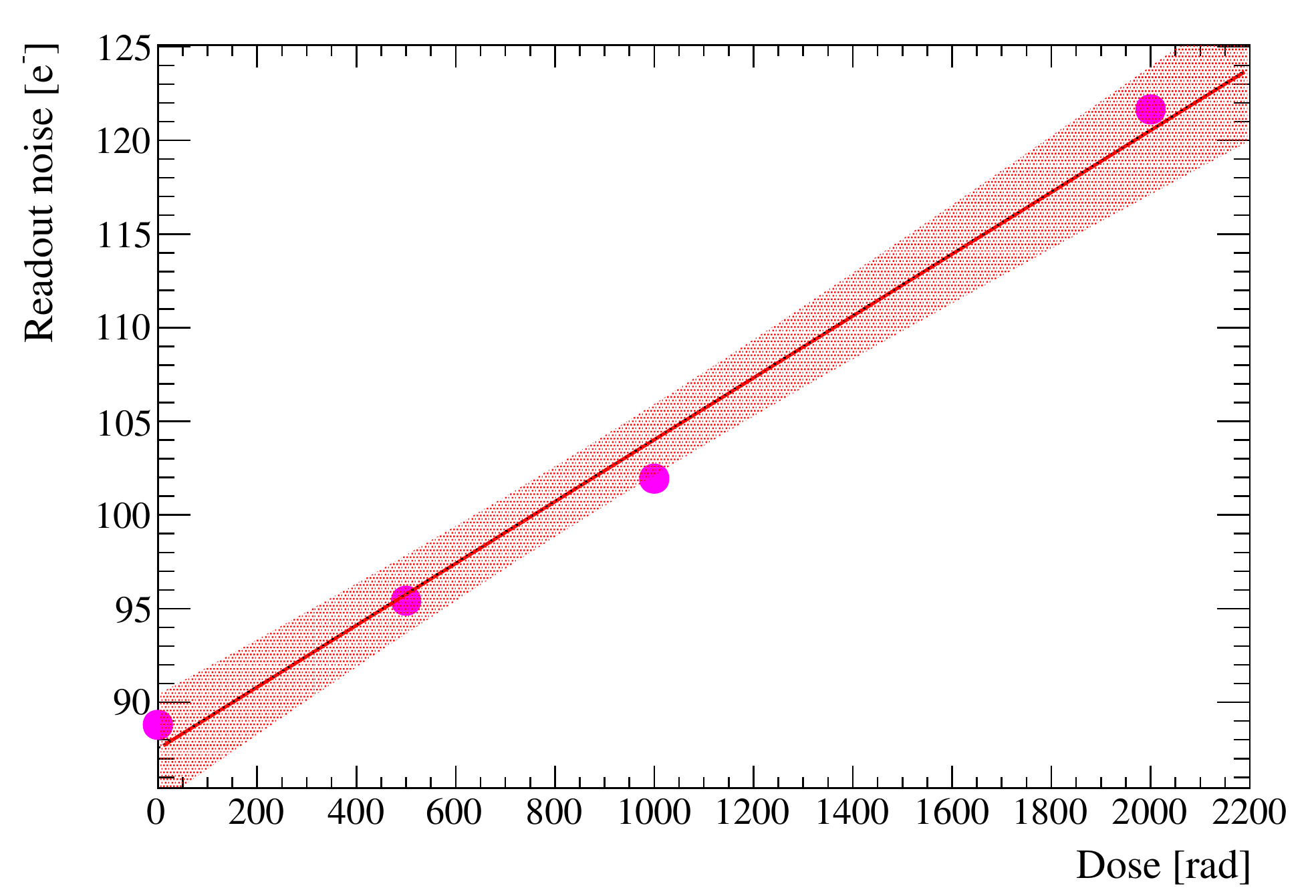}}%
\end{tabular}
\end{center}
\caption[example] 
{ \label{fig:performance1}Degradation of dark current (left panel) and readout noise (right panel). Lines and shadows indicate the best fit linear function and its $95\%$ confidence interval.}
\end{figure}

Dark current is the charge per unit time flowing into the sense node when it is not irradiated with an X-ray. It was derived by linear fitting the relationship between the mean of the pedestal and the integration time. Fig. \ref{fig:performance1} (left panel) shows that the dark current increased in proportion to the radiation dose, with a $239\pm43\%$ increase at 500 rad, which approximately corresponds to the dose in 5 years in orbit. After 2000 rad damage, it increased by $957\pm172\%$.

\begin{figure} [ht]
\begin{center}
\begin{tabular}{c} 
\includegraphics[height=7cm]{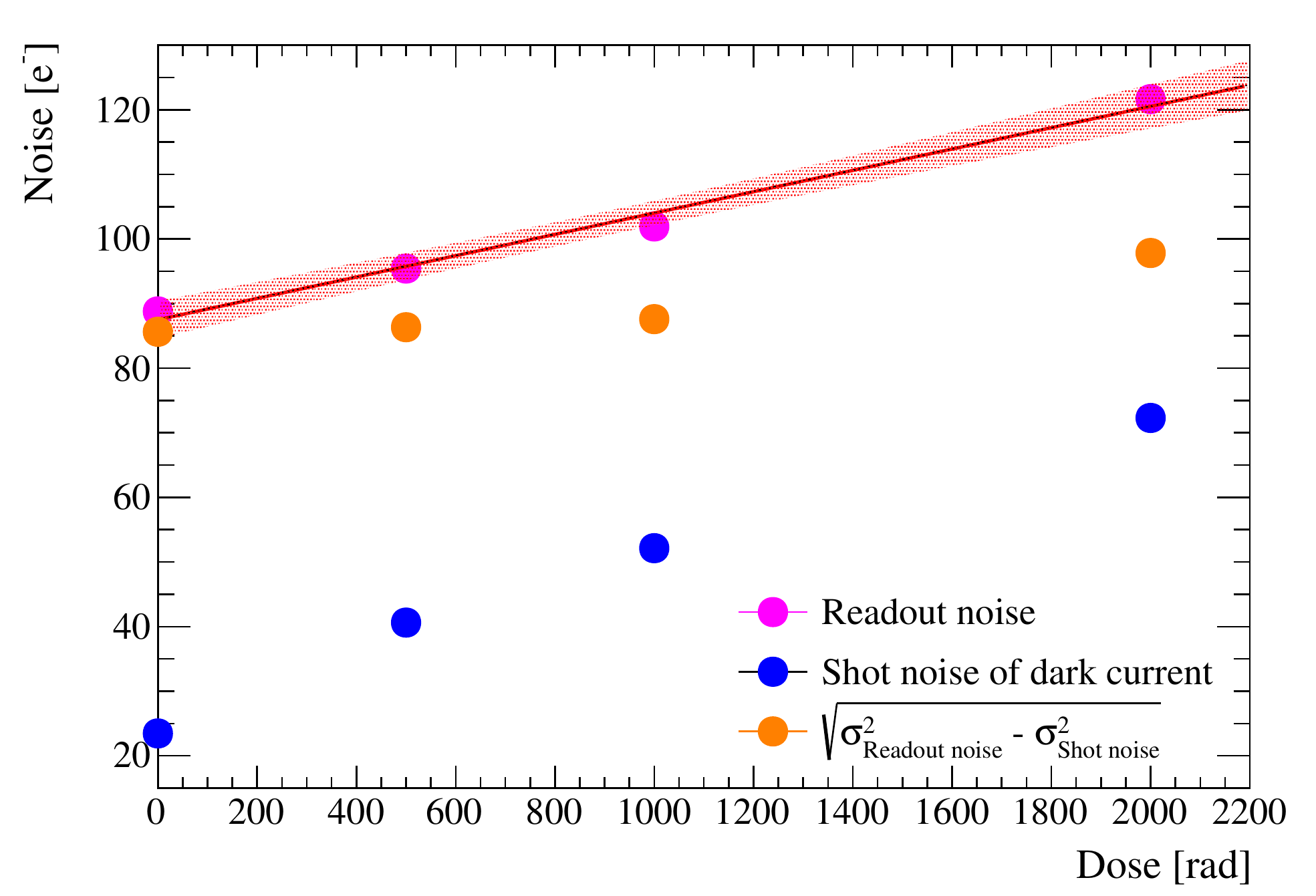}
\end{tabular}
\end{center}
\caption[example] 
{ \label{fig:Noise}
Contributions of the shot noise of dark current and the other noise component to the readout noise.}
\end{figure}

Readout noise is originated from dark current and CMOS circuit. It is expressed as the standard deviation of the Gaussian function fitted to the pedestal. We measured readout noise at an integration time of 10 ms. Fig. \ref{fig:performance1} (right panel) shows that the readout noise increased in proportion to the radiation dose, with a $9\pm1\%$ increase at 500 rad. After 2000 rad irradiation, it increased by $38\pm3\%$. Before the irradiation, the shot noise of the dark current was $\sqrt{55 {\rm ~e^-/ms/pix} \times 10 {\rm ~ms}}=23{\rm ~e^-/pix}$. It is negligible in the readout noise of $90\mathrm{e^-}$. Fig. \ref{fig:Noise} shows readout noise, dark current shot noise, and their residual. After 2000 rad irradiation, the contribution of shot noise to readout noise was about 60$\%$. The increase in readout noise due to irradiation cannot be explained by dark current only. Regarding the origin of readout noise except for dark current, previous studies of SOIPIX suggest increasing the parasitic capacitance of the sense node\cite{HAGINO2020164435} and changing transistor characteristics from the positive charge that accumulates in the BOX layer\cite{HARA2019426}.
 
\begin{figure} [ht]
\begin{center}
\begin{tabular}{c} 
\subfigure{%
\includegraphics[height=5cm]{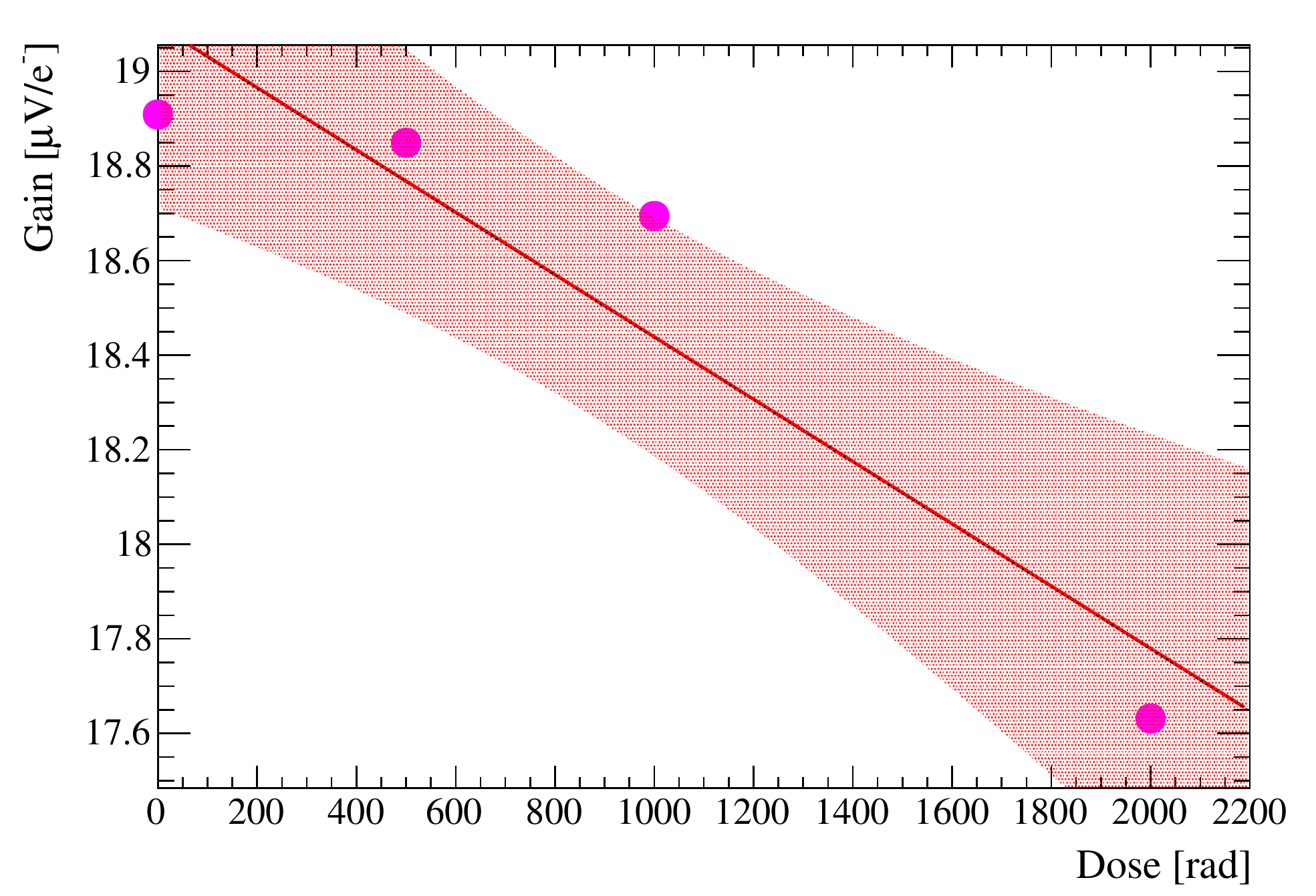}}%
\subfigure{%
\includegraphics[height=5cm]{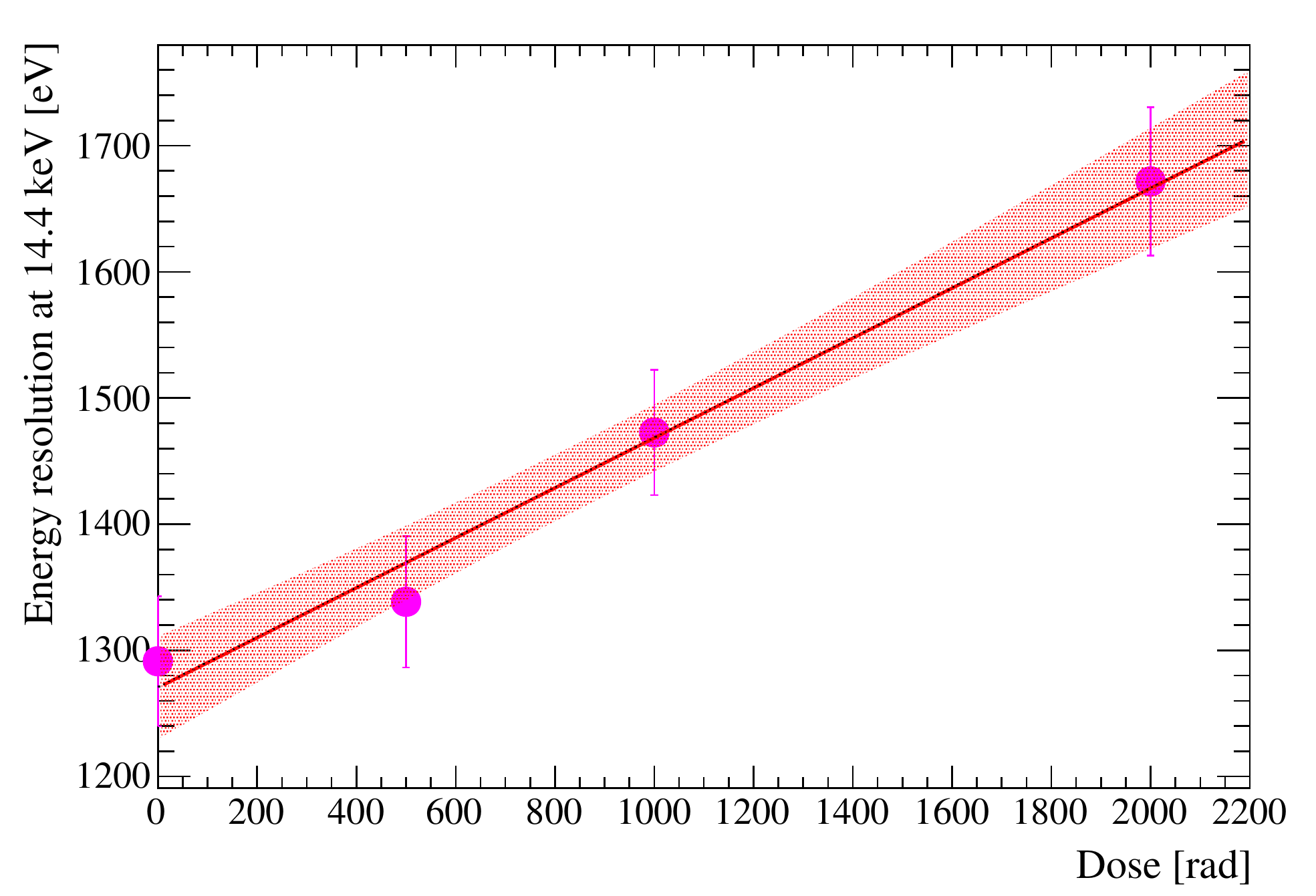}}%
\end{tabular}
\end{center}
\caption[example] 
{ \label{fig:performance2}Degradation of gain (left panel) and energy resolution (right panel). Lines and shadows indicate the best fit linear function and its $95\%$ confidence interval.}
\end{figure}

Gain and energy resolution was evaluated using two radioisotopes, $^{57}$Co and $^{109}$Cd. The emission line was fitted with the Gaussian function. The gain was derived from the relationship between the mean value of the Gaussian and the known energy of X-ray from the isotopes. Fig. \ref{fig:performance2} (left panel) shows that after 500 rad irradiation. The gain decreased by only $1.7\pm0.4\%$. However, the gain was significantly reduced by $6.9\pm1.7\%$ once the dose reached 2000 rad.

\begin{figure} [ht]
\begin{center}
\begin{tabular}{c} 
\includegraphics[height=7cm]{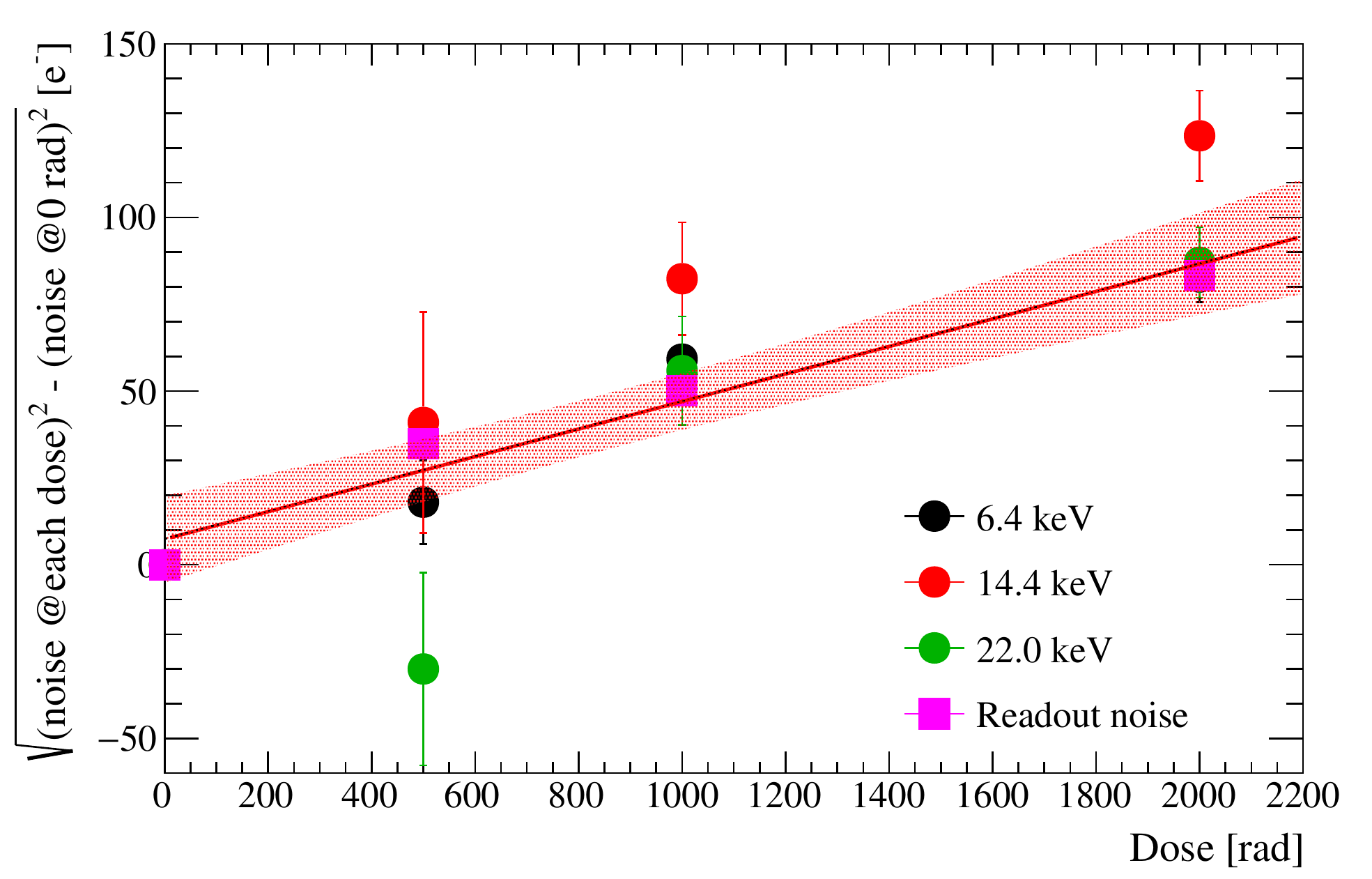}
\end{tabular}
\end{center}
\caption[example] 
{ \label{fig:noise2}
Increase in the readout noise and the energy resolution.}
\end{figure}

The energy resolution was evaluated as Full Width at Half Maximum (FWHM) by fitting the emission lines with a Gaussian function. Fig. \ref{fig:performance2} (right panel) shows that the energy resolution increased in proportion to the dose with an $8\pm1\%$ increase at 500 rad for 14.4 keV X-ray. After 2000 rad dose, it increased by $31\pm3\%$. Fig. \ref{fig:noise2} shows the square root of the squared difference between the readout noise at each dose and that before the irradiation. There is consistency comparing the increase in readout noise indicated by the red confidence region and energy resolution at each energy indicated by the black, red and green dots. Therefore, the increase in readout noise can explain energy resolution degradation.

\section{DISCUSSION}
\begin{figure} [ht]
\begin{center}
\begin{tabular}{c} 
\subfigure{%
\includegraphics[height=5cm]{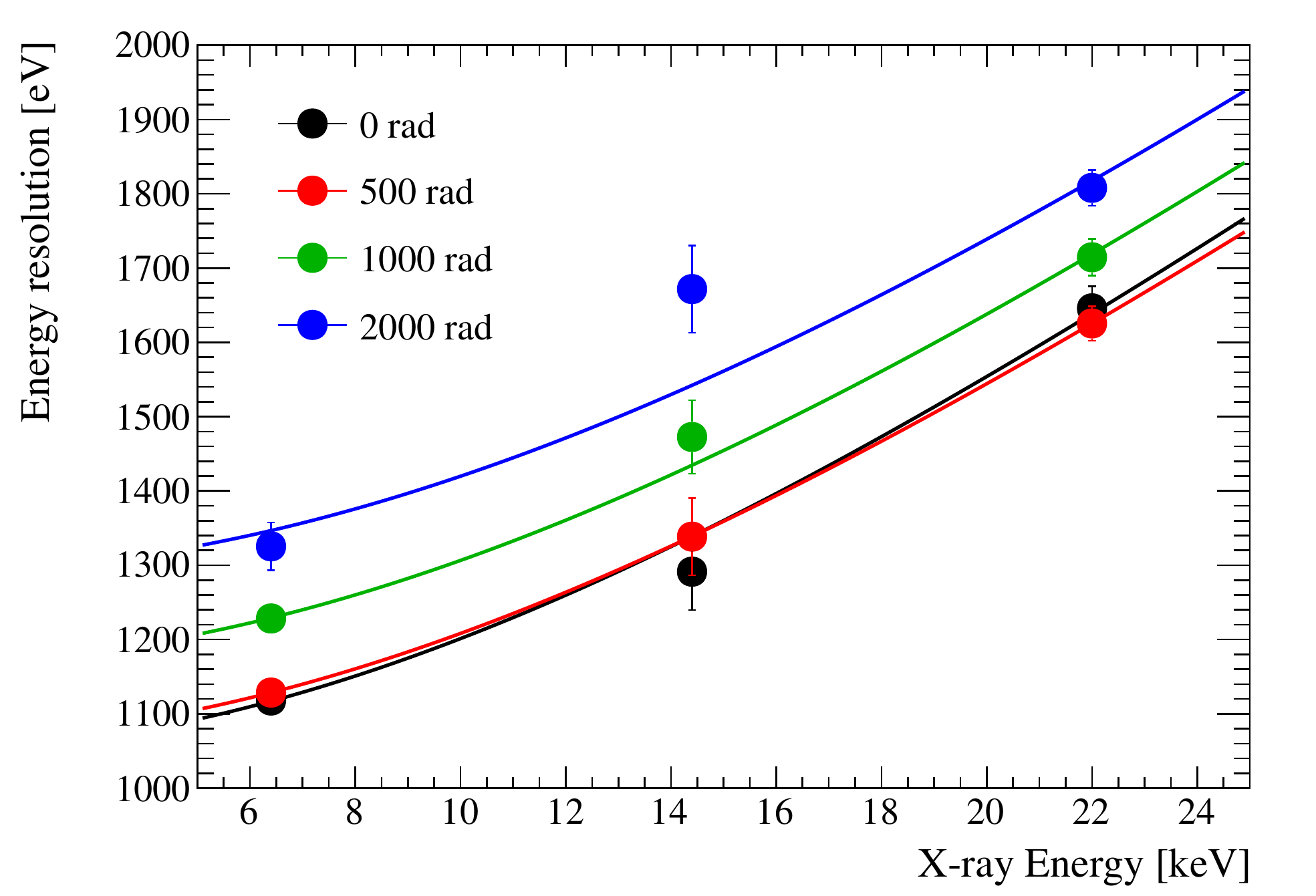}}%
\subfigure{%
\includegraphics[height=5cm]{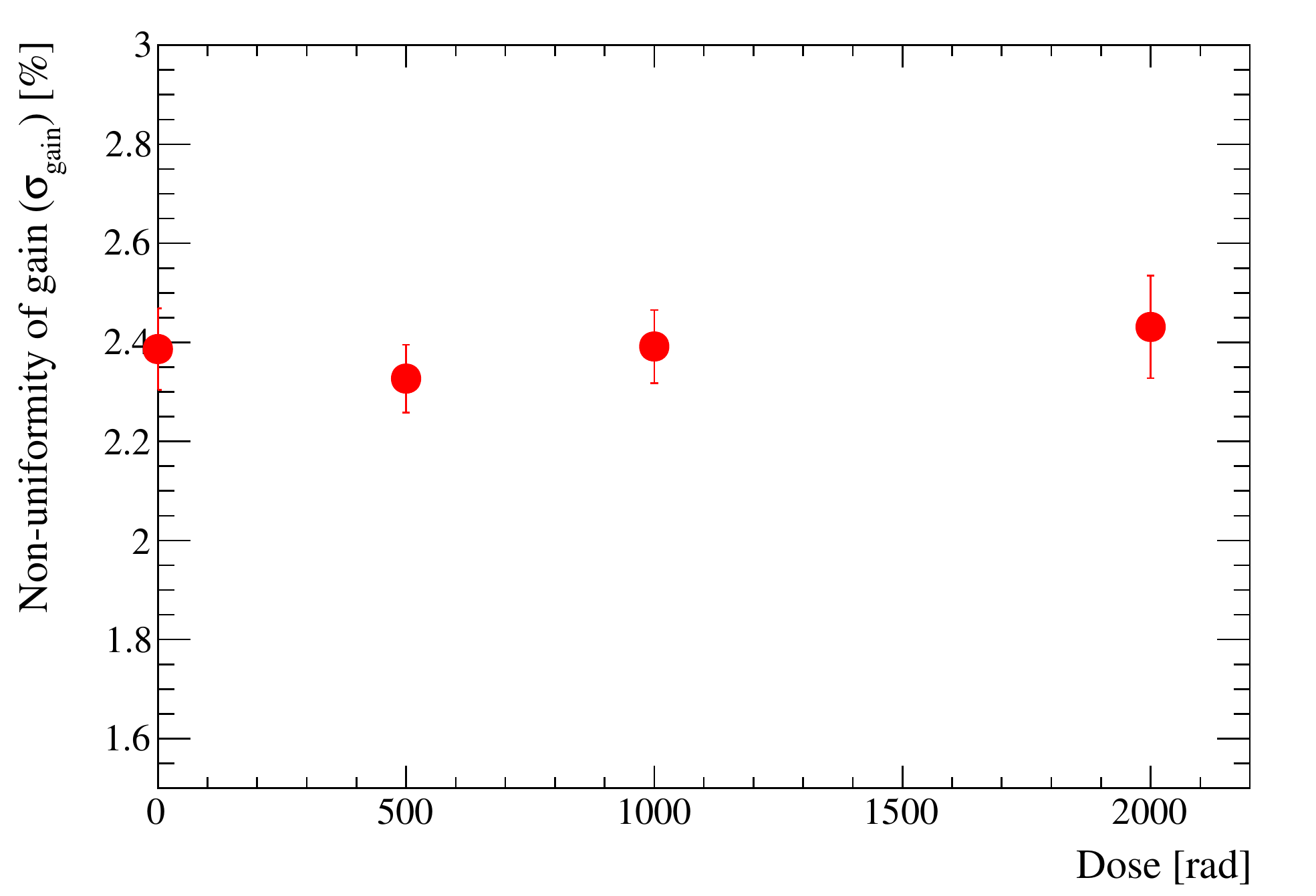}}%
\end{tabular}
\end{center}
\caption[example] 
{ \label{fig:noise_sepa}Energy dependence of the energy resolution at each dose (left panel) and lines show the best fit model function of Eq. \ref{eq:noise}. Non-uniformity of gain between pixels as a function of dose (right panel).}
\end{figure}

We investigated the non-uniformity of readout noise and gain between pixels by irradiating proton with various dose. When observing X-ray polarization, it is necessary to pay attention to a non-uniformity of gain and readout noise in pixels because the signal crosses multiple pixels. Non-uniformity of gain was evaluated using the energy dependence of noise. An Energy resolution can be written as\cite{KODAMA2021164745},

\begin{equation}
\label{eq:noise}
\Delta E = 2.355W\sqrt{\left(\frac{FE}{W}\right)+\sigma^{2}_{\mathrm{inde}}+\left(\frac{\sigma_{\mathrm{gain}} E}{W}\right)^2},
\end{equation}
where $W$ is the mean electron-hole pair creation energy in Si $(3.65~\mathrm{eV/e^{-}})$, $E$ is the X-ray energy, $F$ is the Fano factor (0.12), $\sigma_{\mathrm{inde}}$ is an energy-independent component such as mainly readout noise in the unit of the number of electrons and $\sigma_{\mathrm{gain}}$ is non-uniformity of gain between pixels relative to the mean gain. In general, energy resolution can be divided into two components in terms of energy dependence. These include Fano noise, which is proportional to the square root of energy, and an energy-independent component such as readout noise. When we consider the energy resolution of the spectrum obtained by adding all pixels, in addition to these components, an energy-proportional component due to non-uniformity of gain between pixels contributes. The best fit model function of Eq. \ref{eq:noise} are shown in Fig. \ref{fig:noise_sepa} (left panel). Fig. \ref{fig:noise_sepa} (right panel) shows best fit value of $\sigma_\mathrm{gain}$ as a function of dose. The gain non-uniformity was constant within 0.1$\%$ with 2000 rad irradiation.

\begin{figure} [ht]
\begin{center}
\begin{tabular}{c} 
\subfigure{%
\includegraphics[height=5cm]{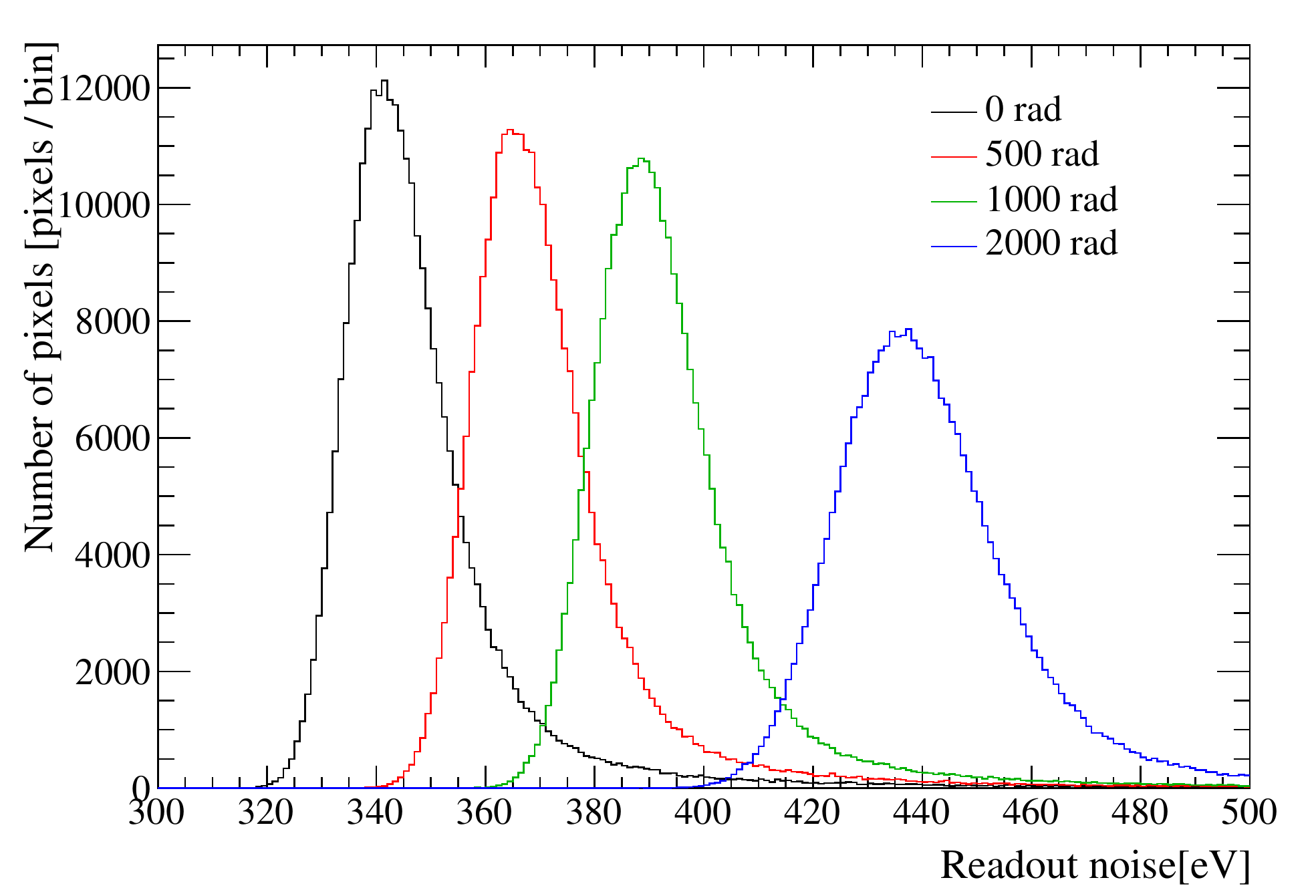}}%
\subfigure{%
\includegraphics[height=5cm]{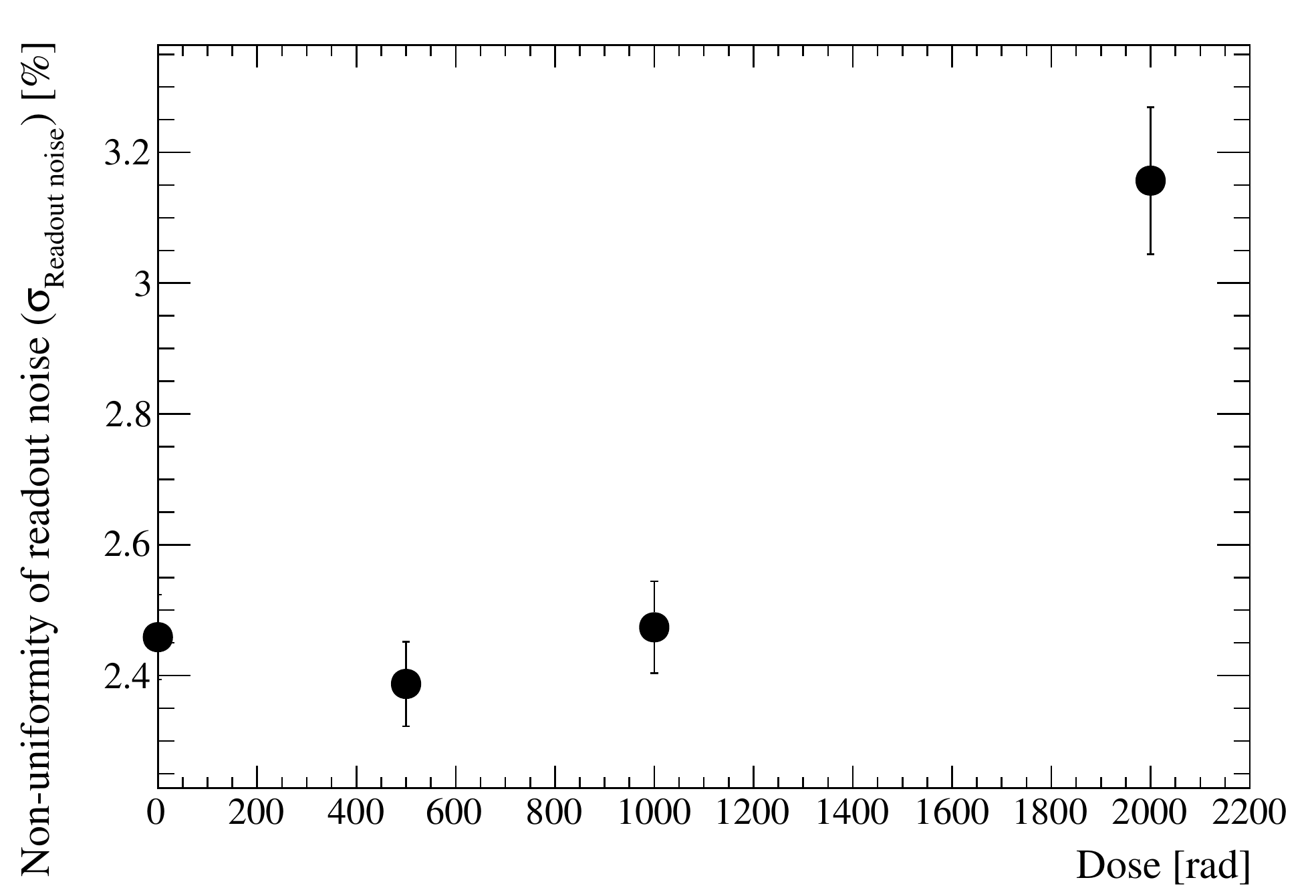}}%
\end{tabular}
\end{center}
\caption[example] 
{ \label{fig:SD}Histogram of readout noise between pixels at each dose (left panel) and non-uniformity of readout noise between pixels as a function of dose (right panel).}
\end{figure}

We also evaluated the non-uniformity of readout noise between pixels. Fig. \ref{fig:SD} (left panel) shows a histogram of readout noise between pixels. The mean and the standard deviation of readout noise increase with dose. 
We derived the mean and the standard deviation by fitting the histogram of Fig. \ref{fig:SD} (left panel) with a Gaussian function. As with non-uniformity of gain, non-uniformity of readout noise ($\sigma_\mathrm{Readout noise}$) was defined as the ratio of the standard deviation to the mean. Fig. \ref{fig:SD} (right panel) shows $\sigma_\mathrm{Readout noise}$ as a function of dose. The readout noise non-uniformity was constant within 0.1$\%$ up to 1000 rad irradiation. At 2000 rad, the non-uniformity of readout noise clearly increased by about 0.5$\%$. This indicates that at 2000 rad, the performance degradation mechanism that affects only the readout noise without affecting the gain is beginning to take effect. The cause of the readout noise degradation may be revealed in the detailed investigation of its results.

\section{SUMMARY}
We evaluated spectral performance at doses of 0, 500, 1000, and 2000 rad. After 500 rad irradiation, which corresponds to about 5 years of dose in orbit, the degradation of the energy resolution at 14.4~keV is less than 10$\%$. After 2000 rad irradiation, the degradation of the energy resolution at 14.4~keV is about 30$\%$. We found that the increase in readout noise can explain energy resolution degradation. After 500 rad irradiation, the non-uniformity of readout noise and gain between pixels is constant within 0.1$\%$. After 2000 rad irradiation, the non-uniformity of readout noise clearly increased by about 0.5$\%$. The results may reveal the cause of the readout noise degradation. From these results, it was found that INTPIX has sufficient radiation tolerance for X-ray polarimetry in space.
\acknowledgments
We acknowledge the valuable advice and manufacture of the INTPIXs by the personnel of LAPIS Semiconductor Co., Ltd. This study was supported by MEXT/JSPS KAKENHI Grant-in-Aid for Scientific Research on Innovative Areas 25109002 (Y.A.) and 25109004 (T.G.T., T.T., K.M., A.T., and T.K.), Grant-in-Aid for Scientific Research (B) 25287042 (T.K.), 22H01269 (T.K.), Grant-in-Aid for Young Scientists (B) 15K17648 (A.T.), Grant-in-Aid for Challenging Exploratory Research 26610047 (T.G.T.), and Grant-in-Aid for Early Career Scientists 19K14742 (A.T.). This study was also supported by the VLSI Design and Education Center (VDEC), Japan, the University of Tokyo, Japan, in collaboration with Cadence Design Systems, Inc., USA, Mentor Graphics, Inc., USA, and Synopsys, Inc, USA.
% References
\nocite{*}
\bibliography{report.bib} % bibliography data in report.bib

\begin{thebibliography}{1}

\bibitem{10.1117/12.2312098}
Tsuru, T.~G., Hayashi, H., Tachibana, K., Harada, S., Uchida, H., Tanaka, T.,
  Arai, Y., Kurachi, I., Mori, K., Takeda, A., Nishioka, Y., Takebayashi, N.,
  Yokoyama, S., Fukuda, K., Kohmura, T., Hagino, K., Ohno, K., Negishi, K.,
  Yarita, K., Kawahito, S., Kagawa, K., Yasutomi, K., Shrestha, S., Nakanishi,
  S., Kamehama, H., and Matsumura, H., ``{Kyoto's event-driven x-ray astronomy
  SOI pixel sensor for the FORCE mission},'' in [{\em High Energy, Optical, and
  Infrared Detectors for Astronomy VIII}{\nolinebreak\hspace{0.1em}]},
  Holland, A.~D. and Beletic, J., eds.,  {\bf 10709},  132 -- 142,
  International Society for Optics and Photonics, SPIE (2018).

\bibitem{Yamada_2018}
Yamada, M., Ono, S., Tsuboyama, T., Arai, Y., Haba, J., Ikegami, Y., Kurachi,
  I., Togawa, M., Mori, T., Aoyagi, W., Endo, S., Hara, K., Honda, S., and
  Sekigawa, D., ``Development of monolithic pixel detector with {SOI}
  technology for the {ILC} vertex detector,'' {\em Journal of
  Instrumentation}~{\bf 13},  C01037--C01037 (jan 2018).

\bibitem{SASAKI2020164426}
Sasaki, T., Shin-ya, M., Mitsui, S., Nishimura, R., Yanagi, K., Miyoshi, T.,
  and Arai, Y., ``X-ray tri-axial stress analysis system using two monolithic
  soi pixel detectors,'' {\em Nuclear Instruments and Methods in Physics
  Research Section A: Accelerators, Spectrometers, Detectors and Associated
  Equipment}~{\bf 979},  164426 (2020).

\bibitem{YARITA2019457}
Yarita, K., Kohmura, T., Hagino, K., Kogiso, T., Oono, K., Negishi, K.,
  Tamasawa, K., Sasaki, A., Yoshiki, S., Tsuru, T.~G., Tanaka, T., Matsumura,
  H., Tachibana, K., Hayashi, H., Harada, S., Takeda, A., Mori, K., Nishioka,
  Y., Takebayashi, N., Yokoyama, S., Fukuda, K., Arai, Y., Miyoshi, T.,
  Kurachi, I., and Hamano, T., ``Proton radiation damage experiment for x-ray
  soi pixel detectors,'' {\em Nuclear Instruments and Methods in Physics
  Research Section A: Accelerators, Spectrometers, Detectors and Associated
  Equipment}~{\bf 924},  457--461 (2019).
\newblock 11th International Hiroshima Symposium on Development and Application
  of Semiconductor Tracking Detectors.

\bibitem{HAGINO2020164435}
Hagino, K., Yarita, K., Negishi, K., Oono, K., Hayashida, M., Kitajima, M.,
  Kohmura, T., Tsuru, T.~G., Tanaka, T., Uchida, H., Kayama, K., Amano, Y.,
  Kodama, R., Takeda, A., Mori, K., Nishioka, Y., Yukumoto, M., Hida, T., Arai,
  Y., Kurachi, I., Hamano, T., and Kitamura, H., ``Radiation damage effects on
  double-soi pixel sensors for x-ray astronomy,'' {\em Nuclear Instruments and
  Methods in Physics Research Section A: Accelerators, Spectrometers, Detectors
  and Associated Equipment}~{\bf 978},  164435 (2020).

\bibitem{HARA2019426}
Hara, K., Aoyagi, W., Sekigawa, D., Iwanami, S., Honda, S., Tsuboyama, T.,
  Arai, Y., Kurachi, I., Miyoshi, T., Yamada, M., and Ikegami, Y., ``Radiation
  hardness of silicon-on-insulator pixel devices,'' {\em Nuclear Instruments
  and Methods in Physics Research Section A: Accelerators, Spectrometers,
  Detectors and Associated Equipment}~{\bf 924},  426--430 (2019).
\newblock 11th International Hiroshima Symposium on Development and Application
  of Semiconductor Tracking Detectors.

\bibitem{KODAMA2021164745}
Kodama, R., Tsuru, T.~G., Tanaka, T., Uchida, H., Kayama, K., Amano, Y.,
  Takeda, A., Mori, K., Nishioka, Y., Yukumoto, M., Hida, T., Arai, Y.,
  Kurachi, I., Kohmura, T., Hagino, K., Hayashida, M., Kitajima, M., Kawahito,
  S., Yasutomi, K., and Kamehama, H., ``Low-energy x-ray performance of soi
  pixel sensors for astronomy, ``xrpix'','' {\em Nuclear Instruments and
  Methods in Physics Research Section A: Accelerators, Spectrometers, Detectors
  and Associated Equipment}~{\bf 986},  164745 (2021).

\end{thebibliography}
\bibliographystyle{spiebib.bst} % makes bibtex use spiebib.bst
\end{document}